\documentclass[twocolumn,a4paper,amsmath,amssymb,aps]{revtex4}

\usepackage{graphicx}
\usepackage{dcolumn}
\usepackage{bm}

\begin{document}



\title[]{Fluctuations of correlations and Green's function reconstruction: \\ role of scattering\\\textit{\small subm. to the Journal of Applied Physics: 2008.01.23, rev.: 2008.03.26, accepted:2008.04.25 .}}

\author{Eric LAROSE}
 \email{eric.larose@ujf-grenoble.fr}
\affiliation{Laboratoire de G\'eophysique Interne et Tectonophysique, Universit\'e J. Fourier, CNRS UMR 5559, BP53, 38041 Grenoble, France.}

\author{Arnaud DERODE}
\affiliation{Laboratoire Ondes et Acoustique, Universit\'e Paris 7, CNRS UMR 7587, ESPCI, Paris, France}

\author{Philippe ROUX}


\author{Michel CAMPILLO}


\begin{abstract}

Correlations of ambient seismic or acoustic vibrations are now widely used to reconstruct the impulse response between two passive receivers as if a source was placed at one of them. This provides the opportunity to do imaging without a source, or \textsl{passive imaging}. Applications include terrestrial and solar seismology, underwater acoustics, and structural health monitoring, to cite only a few. Nevertheless, for a given set of data, correlations do not only yield the Green's function between the sensors. They also contain residual fluctuations that result from an imperfect time or source averaging that might eventually blur the images. In this article, we propose a heuristic model to describe the level of fluctuations of the correlations in the case of non-stationary wavefields, and more particularly in the case of scattering media. The work includes theoretical derivations and numerical simulations. The role of multiple scattering is quantitatively evaluated. The level of fluctuations decreases when the duration and intensity of the diffuse waves increase. The role of absorption is also discussed: absorption is properly retrieved by correlation, but the level of fluctuations is greater, thus degrading the Green's function reconstruction. Discrepancies of our simple model in the case of strong multiple scattering ($k\ell^*\leq 18$) are discussed.
\end{abstract}

\maketitle
\section{Introduction}

Classical waves propagating in heterogeneous media have been subject to increasing interest during the last thirty years. Because diffuse waves show complex waveforms due to the randomness of the medium, they have long been considered to be devoid of any deterministic information. Additionally, it is well established that conventional images (obtained with ultrasounds, radar, seismic waves...) are degraded when scattering increases. Nevertheless, mesoscopic physicists have demonstrated the existence of various wave phenomena that survive, and even develop, in the presence of multiple scattering \cite{akkermans2007,vanalbada1985,wolf1985,derode1995,pine1988,cowan2002}.
This mesoscopic approach has led to an incredible number of applications in optics, acoustics, oceanography \cite{edelmann2005}, and even seismology \cite{larose2004a, margerin2005}. Such applications take advantage of multiple scattering to image, to communicate through, or to monitor heterogeneous media.\\

 Field-field correlation and passive imaging is a more recent idea that strongly benefits from the above developments. The idea is that the correlation of fully diffuse wavefields recorded at two sensors yields the Green's function between them as if one sensor was a source. The connection between correlations and the Green's function is not new and can be derived from the fluctuation-dissipation theorem \cite{kubo1966}. But more recently, Weaver and Lobkis \cite{weaver2001,lobkis2001} proposed another original approach that uses diffuse waves to reconstruct the exact impulse response between two sensors. Their experiments were followed by an amazing breakthrough in seismology \cite{campillo2003,shapiro2005}, where earthquakes are not controlled but sensors easily handled. \\

In practical applications like seismology, we mostly focus our effort on the reconstruction of direct (ballistic) waves within an array of receivers. Such a reconstruction is not trivial to obtain: correlations include the Green's function, plus fluctuations that are not easily washed out. These remnant fluctuations corresponds to the difference between correlations obtained after perfect and after imperfect averaging. These fluctuations, also named \textit{pseudo-noise} in the following manuscript, eventually reduce under time and source averaging. 
 Note that here, the \textit{pseudo-noise} is contained in the correlations and blurs the Green's function, this is different from the \textit{ambient noise} constituted by natural vibrations used as input data for the correlations~\cite{shapiro2004, sabra2005d}. The purpose of the present article is to compare the level of \textit{pseudo-noise} to the level of the perfectly averaged correlation (the \textit{signal}). The \textit{signal} and \textit{pseudo-noise} terminology is chosen here by analogy to active source-sensor experiments. For simplicity, we name signal-to-noise ratio (SNR) the ratio between the level of \textit{signal} in the correlations and the level of \textit{pseudo-noise}.

  In practical applications, estimating the level of \textit{signal} and \textit{pseudo-noise} in the correlations is a central issue. On the one hand, one has to evaluate the minimum amount of data needed to perform some passive images: how many sources to employ, what is the necessary record duration, what distance is best between receivers...? On the other hand, it would be a waste of time to acquire and process an excess of data if the \textit{signal}-to-\textit{pseudo-noise} ratio (SNR) in the correlation is satisfying for less. This SNR quantifies the convergence of the correlations toward the Green's function. Different theories have been developed in helioseismology \cite{gizon2004} and acoustics \cite{weaver2005a, sabra2005c} to describe the convergence with respect to the time of integration $T$, the number of sources $N$, the distance between receivers $r$, the frequency $f$ and bandwidth $\Delta f$ of the recorded signal. 
These theoretical approaches assume stationary wave fields, and are particularly adapted to ambient noise records. What about non-stationary records, like coda waves? Theoretical and experimental works have demonstrated that multiple scattering plays a central role in the time- and space- symmetry of the correlations \cite{vantiggelen2003,malcolm2004,paul2005}. Another point is now to quantify how multiple scattering affects the convergence of correlations. Given a set of sources and receivers, we will see how multiple scattering improves the estimations of the Green's function obtained by correlations. In the present article, we will propose a prediction for the SNR that could be used prior to an experiment. The formulation that we will derive applies to non-stationary wave fields. Our theoretical model will quantitatively describe how an increasing multiple scattering improves the convergence of the correlations.\\


In section~II of the present manuscript, we show an example of Green's function reconstruction in a multiply scattering medium, using finite difference simulations. Sections III and IV are devoted to the evaluation of the \textit{signal} and the \textit{pseudo-noise} in the correlations, theoretical predictions are confronted to numerical simulations. In section V, absorption is added to our theoretical model for the SNR of the correlations. The last section describes the SNR in the diffusion approximation.

\section{Example of correlation and Green's function reconstruction}

First of all, let us begin with a simple illustration of Green's function reconstruction by correlation of diffuse waves.
To simulate wave propagation in heterogeneous open media,  we have chosen to conduct 2-D numerical experiments of acoustic waves~\footnote{The code named ACEL has been developed by M. Tanter, Lab. Ondes \& acoustique (Paris-France). More details on \textsl{http://www.loa.espci.fr/\_michael/fr/acel/aceltest.htm} and in Ref.~\cite{derode2003a}.}. The wave equation is solved by a finite difference simulation (centered scheme), with absorbing boundaries; 
the grid is $50\lambda_0 \times 50 \lambda_0$ large with a $\lambda_0/30$ spatial pitch ($\lambda_0$ is the principal wavelength).\\

\begin{figure}[!htbp]
	\centering
		\includegraphics[width=8cm]{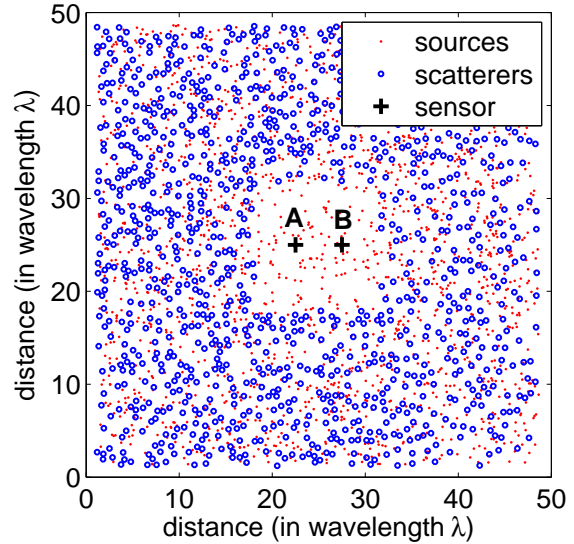}
	\caption{Distribution of the N=1800 sources (dots) and the 1120 scatterers (circles) in the $50\lambda_0 \times 50 \lambda_0$ grid. The number of scatterers varies from 0 to 1200, but the number of sources is kept constant, along with their positions.}
	\label{fig_carte}
\end{figure}

To mimic practical situations like seismology, we have to build an experimental configuration that provides long records (long lasting coda), but presents feeble scattering attenuation between the receivers (labeled A and B) where passive imaging is performed. This would mean a very large grid with low concentrations of scatterers (say $k\ell^{\star}\gg 10$, with $k$ the wave number and $\ell^{\star}$ the transport mean-free path). Since large grids are very time and resource consuming, we chose a configuration with a maximum of scattering in a limited grid. To reduce the effect of scattering attenuation within the array of receivers, we also removed the scatterers from the central region, as can be noticed in Fig.~\ref{fig_carte}. This has the additional virtue of providing an easy to interpret first arrival: the direct wave is simply a wave propagating in a 2D homogeneous medium, whatever the amount of surrounding scatterers (waves reflected on heterogeneities arrive later). 
\begin{table}[!htbp]
		\begin{tabular}{|c|c|c|c|c|}
		\hline
		configuration \#   & 1         & 2     & 3   & 4      \\\hline
			number of scat. & 0         & 400   & 800 & 1200   \\\hline
			$\ell^\star$         & $\infty$  & 5.7   & 2.8 & 1.9      \\\hline
			$k\ell^{\star}$      & $\infty$  & 36    & 18  & 12    \\\hline
			D                    & $\infty$  & 2.8  & 1.4  & 1      \\\hline
			$\tau_D$             & 0         & 34   & 68   & 95    \\\hline
			$\tau_\sigma$             & -         &  110   & 240    & 400     \\\hline

		\end{tabular}
	\caption{physical parameters of the simulations (units in $\lambda_0$, $\lambda_0^2/T_0$ and $T_0$).}
	\label{table}
	
\end{table}

\begin{table}[!htbp]
		\begin{tabular}{|c|c|}
		\hline
		Notation   & Description\\\hline
			$t$ & time (variable) in records $s(t)$\\\hline
			$T_0$      & central period of the records \\\hline
			$T$                    & record duration     \\\hline
	   	$	\tau$         & time lag of the correlations\\\hline
			$\tau_\sigma$             &  decay time of records $s(t)$\\\hline
			$\tau_c$ & coherence time of the diffuse waves \\\hline
			$\tau_a$             & absorption time \\\hline
		\end{tabular}
	\caption{Time notations in the manuscript.}
	\label{table_time}
	
\end{table}

\begin{figure}[!htbp]
	\centering
		\includegraphics[width=8cm]{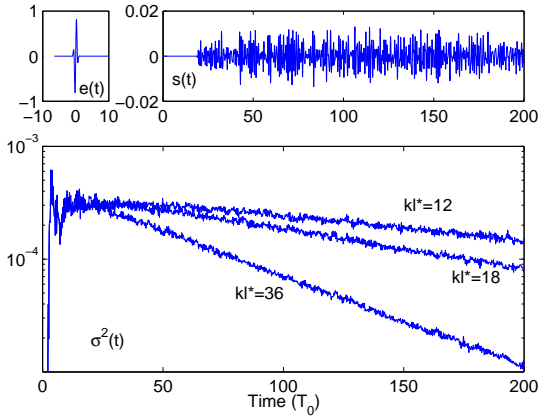}
	\caption{ Top left: source signal $e(t)$. Top right: example of one waveform $s(t)$ received in the multiple scattering medium (experiment number 4, $k\ell^{\star}=12$). Bottom: intensity $\sigma^2$ of the wavefield averaged over 1800 sources, for the three scattering media.}
	\label{waveform}
\end{figure}

 A set of 1800 sources is randomly distributed over the grid and is kept fixed throughout the experiments (see Fig.~\ref{fig_carte}).
The signal $e(t)$ emitted by each source is a pulse with a center frequency $f_0$ and a gaussian envelope (100\% bandwitdh at -6dB). Waveforms $s_{A}(t)$ and $s_{B}(t)$ are recorded at $A$ and $B$ during 200 oscillations $T_0$. The distance $r$ between the two receivers varies from 1 to 10~$\lambda_0$. Typical waveforms $e(t)$ and $s(t)$ are plotted in Fig.~\ref{waveform}. The long tail of the record in Fig.~\ref{waveform}-(b), similar to the seismic coda, corresponds to waves multiply scattered on the surrounding heterogeneities. The exponential decay of the averaged intensity $\sigma^2(t)$ in Fig.~\ref{waveform}-(c) is clearly visible for times greater than 50$T_0$. The decay time is determined by the scattering properties and the absorbing boundary conditions (open medium), and can be fitted by $\sigma(t)=\sigma_0 e^{-t/\tau_\sigma}$. The decay times $\tau_\sigma$ corresponding to the different configurations are reported on Tab.~\ref{table}. All different time notations are also recalled on Tab.~\ref{table_time}. The velocity in the medium is $c=\lambda_0 f_0$. To test the effect of diffusion on the correlations, we have conducted different sets of simulations with different number of scatterers. Each set of simulation is composed of 1800 numerical runs, one for each source (one source at a time). Note that this is different from correlation of ambient vibriations, where sources are continously and simultaneously excited. The scatterers are  empty cavities of diameter $\lambda_0/3$ randomly distributed on the grid (see Fig.~\ref{fig_carte}). Their scattering cross-section was numerically estimated in average over the frequency band of interest: $\Sigma=1.6\lambda_0$, along with their transport cross section: $\Sigma^\star=1.1\lambda_0$. Table~\ref{table} summarizes the physical properties of the simulated media  for the four numerical configurations. This includes the number of scatterers (whose density is $n$), the transport mean free path $\ell^\star=\frac{1}{n\Sigma^{\star}}$, the diffusion constant $D=\frac{c\ell^{\star}}{2}$ and the Thouless time $\tau_D=\frac{R^2}{4D}$ (where $R^2$ is the average of the square of the source-receiver distance).
Note that these quantities are evaluated under the "independent scattering approximation".


\begin{figure} [!htbp]
		\includegraphics[width=8cm]{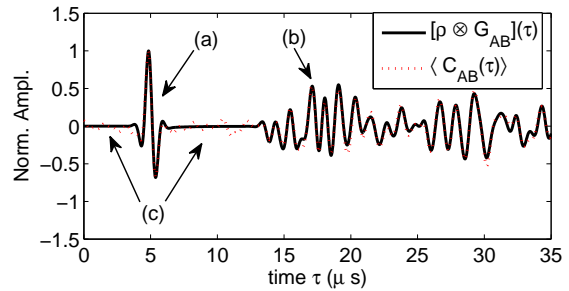}
	\caption{Comparison of the reference waveform (the Green's function) with the averaged correlation $\overline{C_{AB}(\tau)} $ for $r=5\lambda_0$ and for $k\ell^{\star}=12$. (a) denotes the direct wave, (b) the late arrivals corresponding to waves scattered  by surrounding heterogeneities. (c) Weak fluctuations are also visible.}
	\label{fig_correlation}
\end{figure}

The averaged correlation is controlled by three additional independent parameters: the number of sources $N$, the record duration $T$ and the distance $r$ between receivers $A$ and $B$. As an example, we plot in Fig.~\ref{fig_correlation} the correlation averaged over T=200 oscillations and 1800 sources for $r=5\lambda_0$. This correlation is compared to the \textsl{primitive} of the impulse response obtained if $A$ is a source: $e(t)\otimes e(t) \otimes \int G_{AB}(t)dt$ \cite{roux2005a}, where $\otimes$ for convolution. We observe that the full waveform is reconstructed by correlation (a,c). Nevertheless, as the averaging is not perfect, weak fluctuations (b) are also visible, particularly around the direct wave (first arrival). The purpose of the following sections is to estimate the level of the reconstructed Green's function (the \textit{signal}) and the level of fluctuation (the \textit{pseudo-noise}).

\section{Averaged correlation: amplitude of the \textit{signal}}

\subsection{Theory}
 To start with, we consider that a source emits a broadband pulse $e(t)$ that propagates in a heterogeneous and scattering medium and is eventually collected at $A$ and $B$. The record $s(t)$ can be modeled as a non-stationary random signal. Its ensemble average (here, over
all possible source positions) is zero, and $\sigma^2(t)$ denotes its variance. An estimate of $\sigma^2(t)$ can be
obtained by averaging over a large number of sources, as we did on Fig.~\ref{waveform}. If $T$ is the duration of $s(t)$ (the record length for instance), then $\sigma(t)=0$ for  $t>T$ or $t<0$.  Moreover, we assume that $\sigma(t)$, of characteristic decay time $\tau_{\sigma}$, evolves slowly compared to the coherence time of the diffuse waves $\tau_c$ and the central period $T_0$:
$$
\tau_{\sigma} \gg \tau_c , T_0.
$$

 It has been theoretically established that on average over the source position, the correlation of two records yields the Green's function between the receivers \cite{weaver2001,wapenaar2004,roux2005a,colin2006a}. If we note $E\left\{ \right\}$ the averaging over the source position we obtain:

\begin{widetext}
\begin{equation}
E\left\{s_A(t1) s_B(t_2)\right\}=\sigma_A(t_1)\sigma_B(t_2) \rho(t_2-t_1)\otimes \left[\int G^+_{AB}(t_2-t_1) d(t_2-t_1)-\int G^-_{AB}(t_2-t_1) d(t_2-t_1)\right],
\label{eq_expectation}
\end{equation}
\end{widetext}

where $G^+$ and $G^-$ stand for the causal and anti-causal Green's function, and $\rho(\tau)$ the coherence of the diffuse waves. In the simplest approach, $s(t)$ is modeled as a shot noise \textit{i.e.}, a series of replica of the initial pulse $e(t)$ with random and independent arrival times \cite{derode1999}. In that case, it can be shown that $\rho(\tau)$ is simply
\begin{equation}
\rho(\tau)=\frac{\int e(t)e(t+\tau)dt}{\int e^2(t)dt},
\end{equation}
and its typical width $\tau_c=\int\rho^2(t)dt$ is entirely determined by the pulse shape. Nevertheless, depending on the scattering and absorption properties of the medium, $\rho(\tau)$ might be slightly different. In particular, $\rho(\tau)$ is spread when correlation of scatterers are observed in the medium \cite{pnini1989}. This latter point is discussed in section \ref{noise}-A.\\

For clarity, we employ in the following the notation: $\rho(\tau)\otimes \left[\int G^+(\tau)d\tau-\int G^-(\tau)d\tau\right] =\left[\rho \otimes G\right](\tau)$. Due to the spatial symmetry of the configuration and the homogeneity of the source distribution, we have $\sigma_A(t)=\sigma_B(t)=\sigma(t)$.
So far the statistical averages we mentioned where relative to the source position, while
the scatterers positions were fixed. If we replace the source averaging by an average over the
scatterers positions $\left\langle \right\rangle$, we obtain a similar equation, except that the Green's function $\left\langle G\right\rangle_{AB}$ is now the
effective medium Green's function. Correlations would then read:
\begin{widetext}
\begin{equation}
 \left\langle s_A(t1) s_B(t_2)\right\rangle =\sigma_A(t_1)\sigma_B(t_2) \rho(t_2-t_1)\otimes \left[\int \left\langle G\right\rangle^+_{AB}(t_2-t_1)d(t_2-t_1)-\int\left\langle G\right\rangle^-_{AB}(t_2-t_1)d(t_2-t_1) \right ].
\label{eq_expectation2}
\end{equation}
\end{widetext}

 This latter form will not be discussed in the present manuscript, 
in the whole manuscript we do not employ any averaging ovr disorder, but source and/or time averaging. $G^+$ and $G^-$ are therefore the exact Green's functions.

In practical applications like imaging, the repartition of sources and scatterers is fixed.
The Green's function between two sensors A and B is estimated by a double averaging : over a finite
time $T$ and over a finite amount of sources. This estimate writes :
\begin{eqnarray}
 \overline{C_{AB}(\tau)}=\frac{1}{N}\sum_N \int_0^T s_{A}(t)s_{B}(t+\tau)dt\\
=\int_0^T \sigma(t)\sigma(t+\tau)dt \left[\rho \otimes G_{AB}\right](\tau)+F(\tau)
\label{eq_signal}
\end{eqnarray}

The ensemble average (over all possible sources positions) of this estimate is given by the left-hand
side of Eq.~\ref{eq_signal}; it corresponds to the contribution that is useful for the reconstruction of the Green's
function. More quantitatively, it predicts for all times $\tau$ (including late and diffuse arrivals) and
distances $r$ the amplitude of the \textit{signal} part in the correlation.
The right-hand side of Eq.~\ref{eq_signal} corresponds to remnant fluctuations $F$ which are expected to tend to zero
with increasing time $T$ and number of sources $N$. The amplitude of this \textit{pseudo-noise} will be evaluated in
section IV.

\subsection{Numerical validation}

\begin{figure}[!htbp]
	\centering
		\includegraphics[width=8cm]{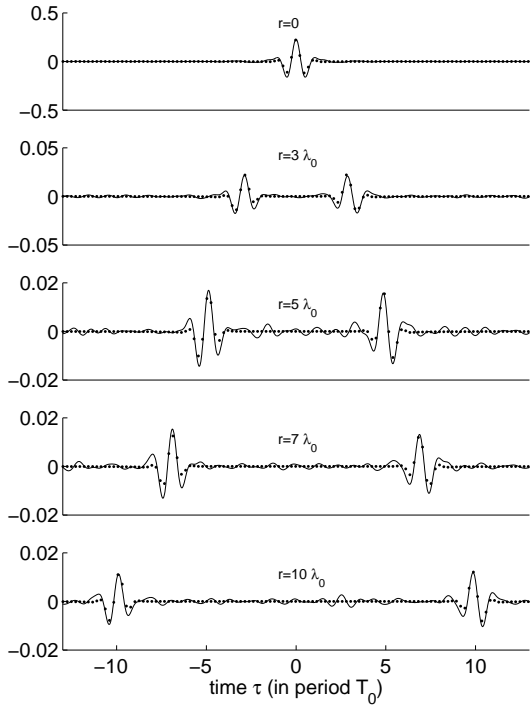}
	\caption{Amplitude of the \textit{signal} in the cross-correlations of diffuse fields. Theory from Eq.~\ref{eq_signal} (dotted line) perfectly fits the correlations (solid lines). Plot $r=5\lambda_0$ corresponds to Fig.~\ref{fig_correlation}.}
	\label{fig:reconstruction_amplitude}
\end{figure}
To confirm quantitatively the level of \textit{signal} in  the correlations (Eq.~\ref{eq_signal}), we plot in Fig.~\ref{fig:reconstruction_amplitude} the correlation $\overline{C_{AB}(\tau)}$ (solid lines) obtained in the numerical simulations (average over $T=200$ oscillations  and 1800 sources), and the theoretical expectation $\int_0^T \sigma(t)\sigma(t+\tau)dt \left[\rho \otimes G_{AB}\right](\tau)$ (dotted lines) for increasing distances $r$. For simplicity, we zoomed into early times $\tau$ where the direct wave is perfectly fitted. Note that reflections arriving later ($\tau\geq 14$) are also reconstructed and fitted with the proper amplitude, though they are not shown here (see the reconstruction of early and late arrival amplitude in Fig.~\ref{fig_correlation}). The agreement is perfect at all distances and all times, meaning that we have a satisfactory theoretical model for the \textit{signal} level in the correlation. \\

To summarize, we have here proposed a model to predict the amplitude of the \textit{signal} in the correlation of diffuse waves. From a practical point of view, in this paper, we are interested at evaluating the SNR of the correlations around the direct wave. Therefore, in the following, the \textit{signal} level will be defined as the maximum of the direct wave, which in theory reads:
\begin{equation}
S_{theo}(r,\tau)=max\left\{\int_0^{ T}\sigma(t)\sigma(t+\tau)dt \left[\rho \otimes G_{AB}\right](\tau)\right\},
\end{equation}

and in the simulated data:
\begin{equation}
S_{num}(r,\tau)=max  \left\{\overline{C_{AB}(\tau)}\right\},
\end{equation}
with the maximum taken  around the first arrival: $\tau\pm r/c$. Note that these definitions can easily extend to any part of the Green's function, including late reflections and coda waves.

\section{Fluctuation of the correlations: amplitude of the \textit{pseudo-noise}.}\label{noise}

\subsection{theory}

Fluctuations of the correlations are visible as long as the averaging is imperfect, and may blur the correlations if not reduced enough. In most applications, we seek to get a fluctuation level as low as possible, but in practice they are rarely negligible. Knowing and predicting the level of fluctuations will allow us to evaluate the relative error in arrival time for application like imaging. It will also allow us to interpret more clearly weak oscillations in the correlations that could either be reflections in the media (real \textit{signal}) or just remnant fluctuations (\textit{pseudo-noise}). The SNR will be defined as the ratio between the amplitude of the averaged correlations, and the level of fluctuations. To begin, we chose to define the \textit{pseudo-noise} in the correlations from their variance:

\begin{equation}
var\left\{\overline{C_{AB}}\right\}=E \left\{\overline{C_{AB}(\tau)}^2\right\}  - E \left\{\overline{C_{AB}(\tau)}\right\}^2,
\end{equation}
              where the bar denotes finite source and time averaging, and the estimate is obtained over source position averaging. The theoretical derivation of the variance is given in appendix A. We assume that 1) $A$ and $B$ are distant by a few wavelengths, 2) as in section II, the decay characteristic time of $\sigma$ is much greater than the propagation time $r/c$, which is itself  greater than the diffuse wave coherence time $\tau_c$.
Then, in the case of a single source, the variance simplifies as :
\begin{equation}
var_{theo} 
 \approx \int_0^T \sigma^2(t)\sigma^2(t+\tau) dt \int \rho^2(t)dt
\label{variance}
\end{equation}

The theoretical SNR for one source can be deduced from above formulas (Eq.~\ref{eq_signal} and Eq.~\ref{variance}):

\begin{widetext}
\begin{align}
SNR_{theo}(r,\tau)=\frac{\left[\rho \otimes G_{AB}\right](\tau)}{\sqrt{\int\rho^2(t)dt}}\times \frac{\int_0^{ T} \sigma(t)\sigma(t+\tau)dt}{\sqrt{\int_0^{T} \sigma^2(t)\sigma^2(t+\tau)dt}}
\label{SNR}
\end{align}
\end{widetext}

This equation can be generalized to $N$ uncorrelated sources, we have:
\begin{equation}
SNR_{theo}(N)=SNR_{theo}(N=1)\sqrt{N} .\label{SNRN}
\end{equation}

The $\sqrt{N}$ dependency is expected even in the case of correlated sources, as long as the correlation
is short-range. From this general form, it is interesting to note that The SNR depends directly on the Green's function $G_{AB}$, and that it is hardly possible to predict the SNR without having an approximative idea of the Green's function. From Eqs.~\ref{SNR} and \ref{SNRN}, we can also deduce that: 
\begin{enumerate}
\item The convergence of the correlation strongly depends on the envelope $\sigma(t)$ of the raw records $s(t)$. Since diffusion strengthens late arrivals, the stronger diffusion the better the SNR. This is a central result of our paper. 
\item The SNR depends on the amplitude of the reference Green's function ($G_{AB}$), so shorter distances $AB$ and shorter times $\tau$ are more easily reconstructed.
\item The broader the spectrum (the smaller $\tau_c$), the better the SNR.
\item The more sources, the better the SNR.
\end{enumerate}
 All these features have been observed in previous experiments \cite{campillo2003,larose2004b, paul2005, weaver2005a}. Moreover, these latter equations can be simplified as follows. First, we assume that the coherence time of the scattered waves $\tau_c=\int\rho^2(t)dt$ is simply determined by the
duration in the initial signal $e(t)$ (shot noise approach). Second, we assume  that the record time $T$ is greater than or of the order of $\tau_\sigma$ (the characteristic decay time of $\sigma$), so that $\sigma(t)$ takes the simple form of $\sigma_0e^{-t/\tau_\sigma}$. The SNR for our simulation configuration now rewrites:

\begin{equation}
SNR_{theo}(r,\tau)\approx\left[\rho \otimes G_{AB}\right](\tau)\sqrt{\frac{\tau_\sigma N}{\tau_c}}
\end{equation}

This result holding for non-stationary wave fields compares to previous results in the case of correlation of ambient stationary noise (see for instance the statistical approach in ~\cite{gizon2004,sabra2005c,weaver2005a} and a geometrical approach in \cite{roux2004,larose2006e}): if we assume a stationary wave field 
 ($\sigma(t)=\sigma_0$) in a rather homogeneous medium, and if we approximate the coherence time by the inverse of the frequency content of the source $\tau_c \propto\frac{1}{\Delta f}$, then we recover the previous prediction of the SNR for the direct wave in 2D:
\begin{equation}
SNR_{theo}(\tau,r)\propto\sqrt{\frac{ T\Delta f c}{rf}},
\end{equation}
Here follows a validation of our model with numerical simulations in the case of non-stationary diffuse wave fields.

\subsection{Numerical validation for $N$=1 source, $T$ variable.}
\begin{figure}[!htbp]
	\centering
		\includegraphics[width=8cm]{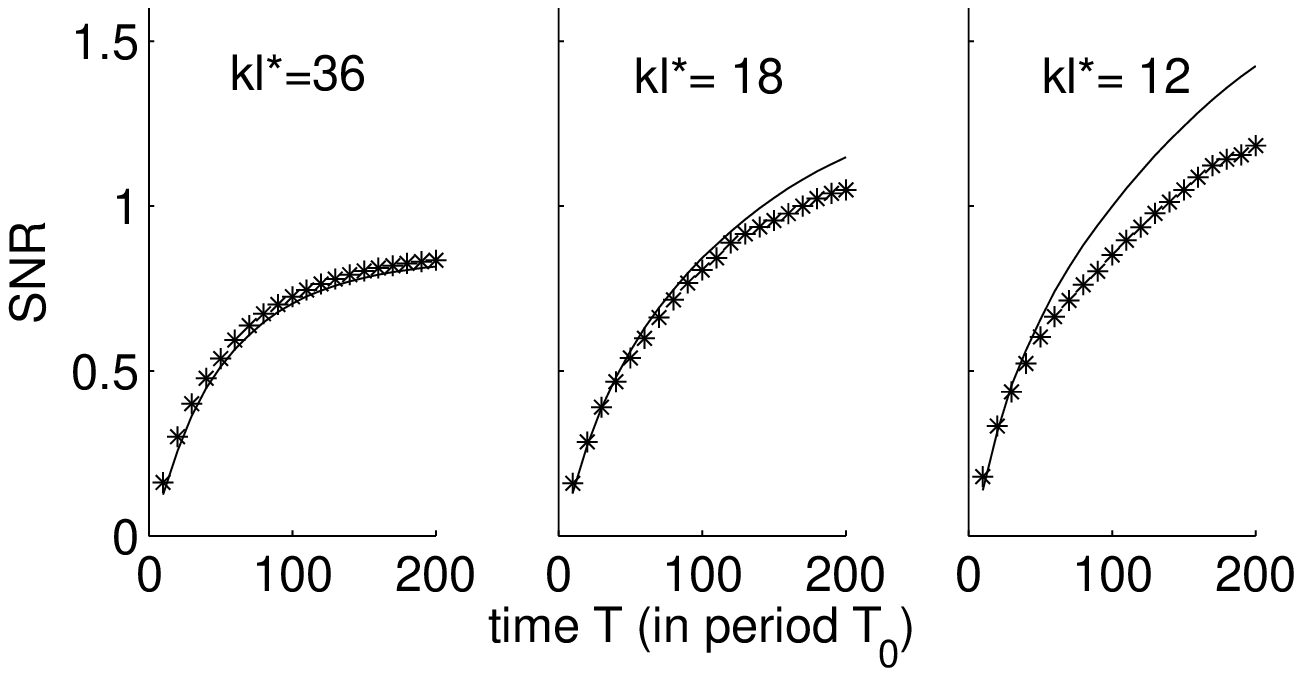}\\
		\includegraphics[width=8cm]{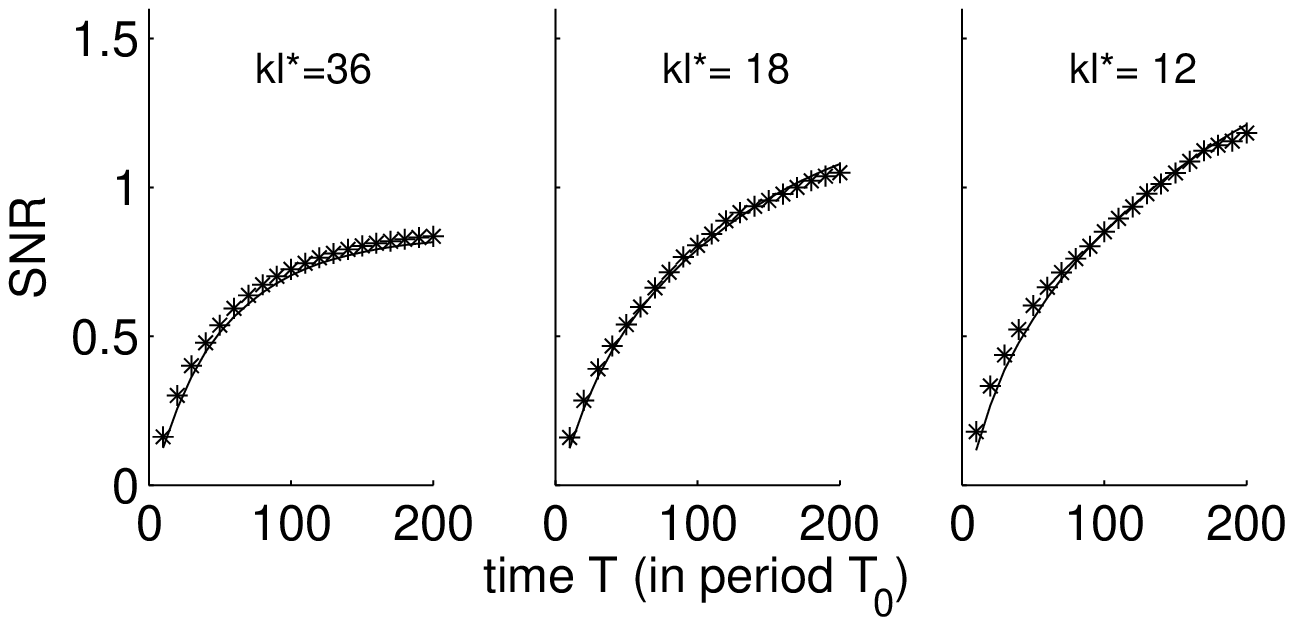}\\
	\caption{Numerical (stars) and theoretical (line) SNR versus the integration time T for N=1 source, and $r=5\lambda_0$. The more the diffusion, the faster the convergence. Top: theory perfectly fits the data for $k\ell^{\star}=36$ (left), but slightly differs from simulations  for large record time and strong diffusion ($\tau_c=0.44$). Bottom: a better fit is obtained by adjusting the coherence time:  $\tau_c=$0.44, 0.50 and 0.61, respectively.}
	\label{fig:SNR_time}
\end{figure}

For practical reasons, we chose to define the \textit{pseudo-noise} level  as the average of the variance of the correlations 
before the direct arrival, and where the Green's function is null. This variance corresponds to the intensity of the fluctuations around $\tau=0$:
\begin{equation}
 var_{num} \left\{C_{AB}\right\}= var_{num} \left\{F\right\}= \overline{C_{AB}^2(\tau\approx 0)}
\end{equation} 
%

with the time average performed over ${-\frac{r}{c} +\tau_c< \tau < \frac{r}{c}-\tau_c}$.
The signal-to-noise ratio is eventually evaluated by the ratio between the \textit{signal} level of the correlation, and this averaged variance:

\begin{equation}
SNR_{theo}=\frac{S_{theo}}{\sqrt{var_{theo}}}\qquad SNR_{num}=\frac{S_{num}}{\sqrt{var_{num}}}.
\end{equation}

As a result, $SNR\leq 1$ when fluctuations dominate, and $SNR\gg 1$ when correlations have converged to the reference Green's function.
In Fig.~\ref{fig:SNR_time}, we compare the theoretical and numerical SNR for $N=1$ source, for an increasing record length $T$, and for three different scattering media. First of all, the SNR increases with the record length $T$. Moreover, the fit between numerical simulations and theoretical SNR is satisfying for $k\ell^{\star}=36$. Nevertheless, for stronger diffusion ($k\ell^{\star}\leq18$) and for large record times (much greater than the scattering mean free time), the prediction for the SNR is found to be slightly inappropriate: the \textit{pseudo-noise} level around $\tau=0$ is found to be stronger than expected. This is clear evidence that our naive model of shot noise for diffuse waves can not account for the complexity of the field after several orders of scattering. In other words, we suspect that there exists some remnant time-correlations in $s(t)$ that appear when scattering is increased. This phenomenon was observed in time-reversal experiments where a saturation of the SNR was observed when scattering was strongly increased~\cite{derode2000}. In our case, we imagine two possible interpretations for this phenomenon. First, the scattered wavefield can excite the same scatterer placed perpendicularly with respect to $A$ and $B$ several times, which results in coherent correlations around $\tau=0$. As the scattering medium does not move (no averaging over disorder), this contribution hardly vanishes under time averaging. Second, there exist recurrent scattering (closed loops) in the medium that can be excited several times by the same source. Both interpretations mean that $s(t)$ can not be modeled as a shot noise with a constant
coherent time. Correlations between the arrival times (scattering paths) induce an
increase of duration of $\rho(\tau)$, which also increases $\tau_c$. Fits of the numerical result yield the following values :  for $k\ell^{\star}=18$ the best fit is obtained for $\tau_c=0.50$, and for $k\ell^{\star}=12$: $\tau_c=0.61$.

Moreover, we clearly see from Fig.~\ref{fig:SNR_time} that the SNR for one source and 200 oscillations is lower or of the order of one, which means that correlations have not (yet) converged: in these conditions the direct wave between $A$ and $B$ for $r=5\lambda_0$ is hardly reconstructed. Additional averaging over sources is needed. This is addressed in the next subsection.

\begin{figure}[!htbp]
	\centering
		\includegraphics[width=4cm]{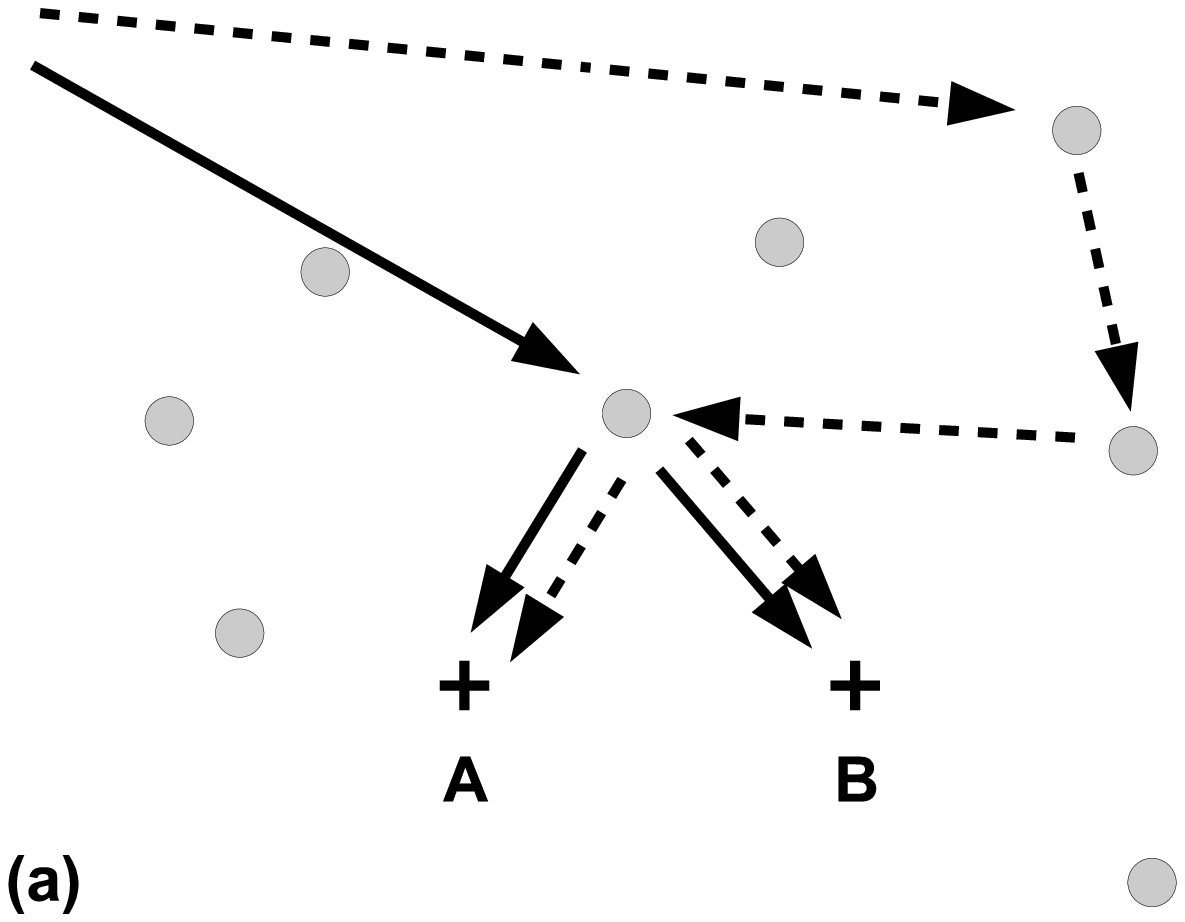}
		\includegraphics[width=4cm]{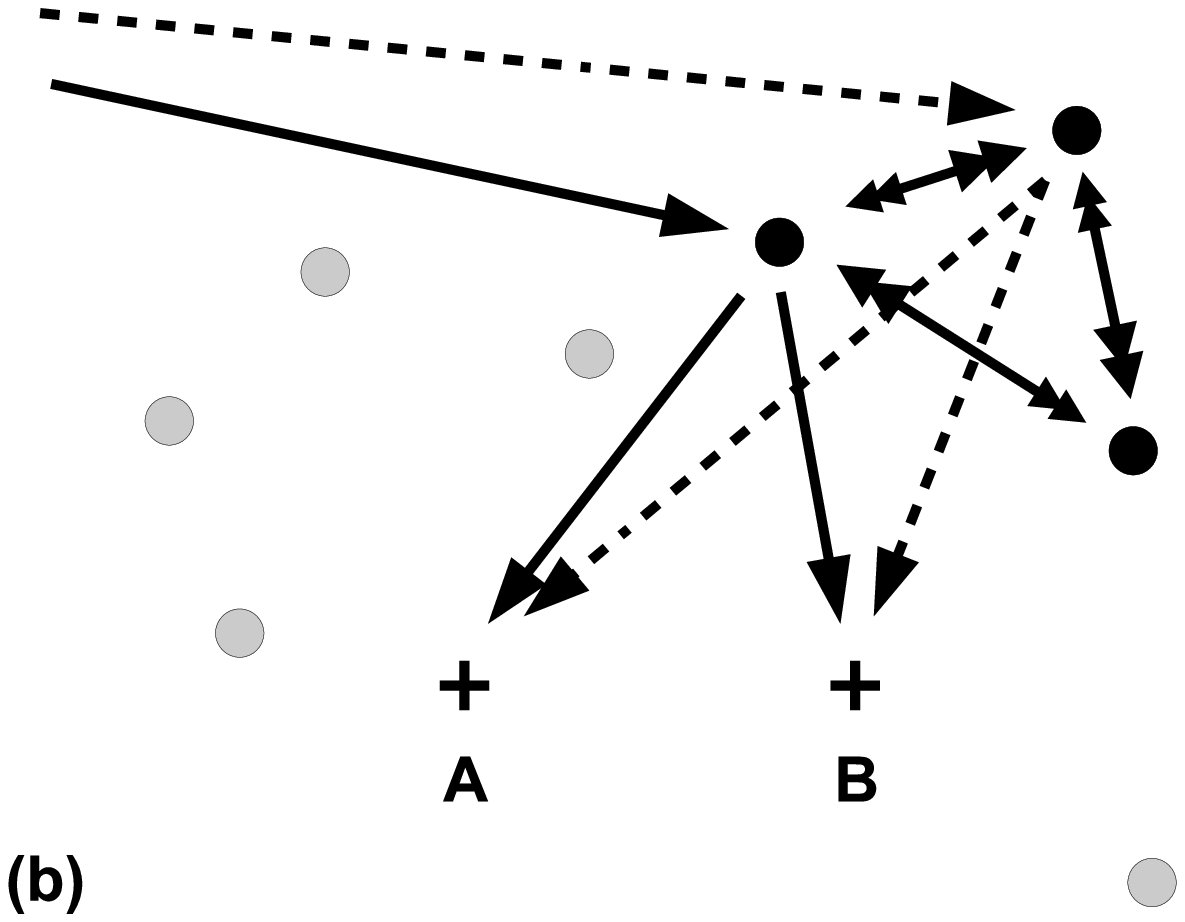}
	\caption{Two possible configurations of paths that result in time-correlation in records $s(t)$. Solid and dashed arrows stand for waves at different time. (a) The same scatterer is excited several times. (b) Existence of recurrent scattering (double arrow and black scatterers).}
	\label{saturation_time}
\end{figure}

\subsection{Numerical validation for T=200 oscillations, N variable.}

\begin{figure}[!htbp]
	\centering
		\includegraphics[width=8cm]{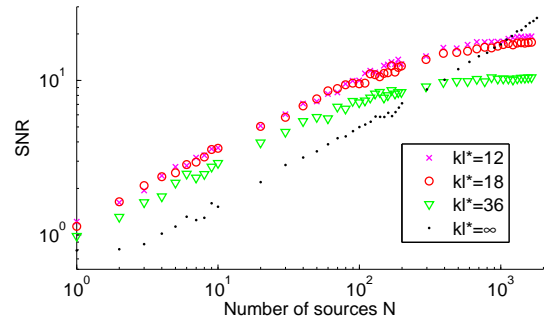}
	\caption{SNR for $r=5\lambda_0$ estimated for an increasing number $N$ of sources and for increasing scattering in the medium for various $k\ell^{\star}$ ranging from $\infty$ (homogeneous medium) to 13 (strong multiple scattering). The stronger the diffusion, the better the correlations. For $N>100$, a discrepancy is observed and might be due to long-range correlation in the disordered medium.}
	\label{fig_SNR_N_diffus}
\end{figure}

We now evaluate the effect of increasing the number of sources on the SNR. In order to have comparable records (with similar statistical content), 1) sources are placed at random (the position of the sources are spatially uncorrelated); 2) each couple of record $s_{A,B}(t)$ originating from the same source is normalized by the maximum of $\left\{\sqrt{\int s_A^2(t)dt},\sqrt{\int s_B^2(t)dt}\right\}$. As diffuse waves have a  coherence length of the order of half the wavelength, assumption 1) is likely to be slightly inappropriate here. 
The residual spatial correlation between the sources' positions has been evaluated in appendix~\ref{Neff}, and its effect is found to be small for the 1800 sources of our numerical experiments.\\

We plot in Fig.~\ref{fig_SNR_N_diffus} the SNR versus the amount of sources used in the averaging, for four different experimental configurations (see Tab.~\ref{table}). After 1800 sources, we obtain a satisfying SNR ($\geq 10$) in all cases. Up to 100 sources, the SNR grows like $\sqrt{N}$,  which means that records can be considered as uncorrelated. As for
one source, the SNR is always better when scattering is stronger (shorter $k\ell^{\star}$). It is particularly interesting to note that all scattering media provide better SNR than the one obtained in the homogeneous one (without scatterers). However, for more than 100 sources, the SNR is no longer a power law. The rate of convergence with the number of sources is slower than expected, which means that the contributions from different sources are not totally independent. As a simple picture, we can again invoke that long range correlations might exist in the scattering media. As a result, the SNR in homogeneous medium eventually goes beyond the SNR with diffuse waves for more than 400 to 900 sources, depending on the amount of scatterers. This is a clear indication that the fluctuation-dissipation theorem does not totally apply to the reconstruction of the Green's function by correlation of diffuse waves. In other words, time averaging, source averaging or ensemble averaging can not be simply interchanged in the case of strong multiple scattering. 



%
%

\section{Role of absorption}

\begin{figure}[!htbp]
	\centering
		\includegraphics[width=8cm]{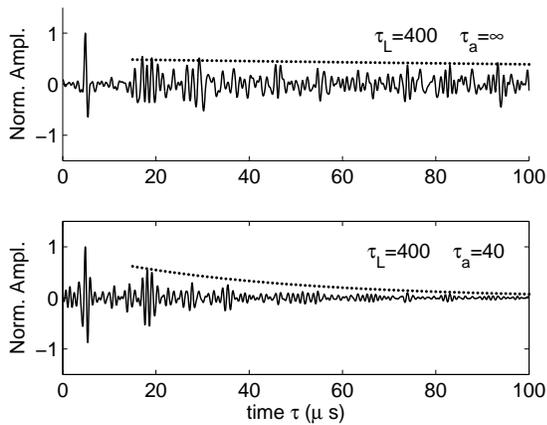}
	\caption{Solid line: correlation without (top) and with (bottom) absorption. Dotted line: theoretical decay. The upper plot is equivalent Fig.~\ref{fig_correlation}. Both amplitude and phase are reconstructed, even when absorption is present.}
	\label{fig:correlation_abs}
\end{figure}

\begin{figure}[!htbp]
	\centering
		\includegraphics[width=8cm]{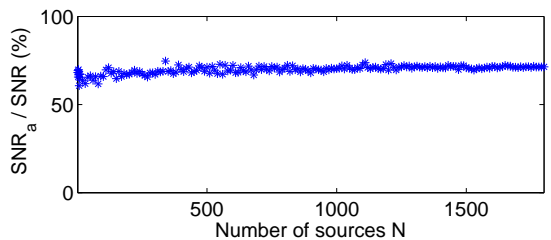}
	\caption{$SNR_a$ to $SNR$ ratio, evaluated for $r=5\lambda_0$, $k\ell^{\star}=36$ and Absorption time here is $\tau_a=40$}
	\label{fig:SNR_abs}
\end{figure}

In practical applications, the medium is often (at least slightly) absorbing. Several authors have worked out the effect of absorption in the correlation\cite{roux2005a,snieder2007,colin2006a}. They proved that in lossy environment, the Green's function (including absorption) is still retrieved by correlation. Nevertheless the role of absorption on the convergence of correlations is not established. We propose in the following to adapt the theory presented in previous sections to the case of weakly absorbing media.\\

First of all, let us consider that the Green's function with absorption $G^a$ is simply connected to the lossless Green's function $G$ as $G^a_{AB}(\tau)=G_{AB}(\tau)e^{-\tau/\tau_{a}}$ with $\tau_{a}$ the absorption time. In our simple model, absorption does not depend on the frequency. The variance of the field is also affected: $\sigma_a(t)=\sigma(t) e^{-t /\tau_{a}}$. First we want to check that the Green's function $G^a$ is actually retrieved by correlation in the numerical simulations. To that end, the same numerical simulations presented above are slightly modified to account for the absorption. Data are processed again, an example of averaged correlations is presented in Fig.~\ref{fig:correlation_abs}. As expected, the reference Green's function $\left[\rho \otimes G^a\right](\tau)$ is retrieved both in phase and amplitude, meaning that the absorption is actually reconstructed in the correlation. Nevertheless, fluctuations are now stronger than in a medium free of absorption. \\

The effect of absorption on the level of fluctuations and the SNR is quantified now. Theory from Eq.~\ref{SNR}\&\ref{SNRN} can be adapted to take absorption into account. For simplicity, we assume that $\sigma_a(t)=\sigma_0 e^{-t\left(\frac{1}{\tau_\sigma}+\frac{1}{\tau_a}\right)}$, and again $T\gg \tau_{\sigma}$ and $\tau_a$. Then

\begin{equation}
\frac{SNR_a}{SNR}= \sqrt{\frac{\tau_a}{\tau_{\sigma}+\tau_a}}e^{-\frac{r}{c\tau_a}}.
\label{abs}
\end{equation}

This formulation contains two contribution. First, absorption reduces the effective length of the coda record. Second, the direct wave is attenuated by a factor $e^{-\frac{r}{c\tau_a}}$. As a result, the SNR decreases with increasing absorption. In other words, the stronger the absorption the lower the SNR.  In Fig.~\ref{fig:SNR_abs}, we plot the ratio $SNR_a/SNR$ around the direct wave versus the number of sources $N$, for the experimental configuration number two (see Tab.~\ref{table}). For the physical values retained here ($r=5\lambda_0$, $\tau_{\sigma}=110$ and $\tau_a=40$), theory from Eq.~\ref{abs} predicts a ratio of SNR of 45\%, which is of the order of experimental results (the average value is 60\%). The discrepancy comes from the fact that the actual form of $\sigma$ is not a simple exponential decay.

\section{Extrapolation to fully diffusive media}
All the above numerical results have been obtained in a configuration where the direct wave within the array of receivers does not suffer from scattering attenuation: no scatterers were placed in the central area of the simulation. In real experiments, scatterers are likely to be distributed everywhere in the medium. We propose in the following to extend our model to this latter configuration, under the diffusion approximation. In this case, the average envelope of the wave field received at a distance $R$ in an infinite medium is:
$$
\sigma(R,t)=\sqrt{\frac{1}{4Dt}e^{-\frac{R^2}{4Dt}}}.
$$
This model is particularly valid when sources are far away from the receivers, meaning $R\gg\ell$. We additionally assume that the distance $r$ between $A$ and $B$ is much smaller than  $R$ so that the envelope is the same for $A$ and $B$. The effective time length of one record is similar to the  Thouless time:
$$
\tau_D= \frac{R^2}{4D},
$$

therefore we would expect the SNR to grow like:
$$
SNR\propto G_{AB} (r,\tau) \sqrt{\frac{N\tau_D}{\tau_c}}.
$$

On the one hand, scattering increases the Thouless time $\tau_D$, thus increasing the SNR. On the other hand, the direct wave is attenuated by scattering. These two effects are in competition. For the direct wave, we can approximate the attenuation by the product of the geometrical spreading and the diffusion attenuation. For simplicity, we also assume that $\ell=\ell^{\star}$. The corresponding SNR then writes:

$$
SNR\propto  \sqrt{\frac{NR^2}{\tau_c c r}\frac{e^{-r/\ell^{\star}}}{\ell^{\star}}},
$$
which formally means that the SNR increases with increasing scattering if $r<\ell^{\star}$. Again, this formal result is a central point of our article. Note that according to this equation, the SNR saturates, then decreases for strong scattering or large distances ($r > \ell^{\star}$).
This is due to the scattering attenuation of the reconstructed Green's function (in the correlation). The best SNR is therefor expected for $r=\ell^{\star}$.

\section{Discussions and amplitude reconstruction}

In the present article, we correlated diffuse waves to reconstruct the Green's function between passive sensors. This processing was carried out without any transformation of the raw record $s(t)$. The only modification that we realized  was normalising each couple $\left\{s_A,s_B\right\}$ by the maximum of $\left\{\int s_A^2(t)dt,\int s_B^2(t)dt\right\}$. In particular, we did not change the frequency content of the records (no whitening), nor change the amplitudes of particular waveforms within $s(t)$. This has the virtue of reconstructing the Green's function both in phase and amplitude. Practically speaking, this allowed us to retrieve the geometrical spreading of the wave, along with the absorption, and confirms the possibility of mapping the attenuation of the medium under investigation "without a source". 
Nevertheless, in most practical applications, the amplitude of the raw records $s(t)$ is strongly modified. Several non-linear transformations have been proposed in the literature: 1-bit processing (retain only the sign)~\cite{larose2004b}, adaptive gain constant (dynamic renormalization that compensate the coda decay), clipping~\cite{sabra2005a}, to cite only a few. In our model, this would be more or less equivalent to compensate the decay of $\sigma$, thus increasing the SNR. On the one hand, these latter processing were shown to greatly improve the final tomographic images: they are particularly adapted when the phase (arrival time) of the wave transports the quantity of interest, like in tomography. On the other hand, these processing are likely to degrade the reconstruction of the amplitude, as it was observed in more recent experiments \cite{larose2007,gouedard2008}. We therefore strongly suggest that pre-processing treatments like 1-bit or whitening be used only when  the  reconstruction of the phase is targeted.

\section{Conclusion}

To conclude this article, we have reported on the convergence of correlations toward the Green's function in the case of independent sources in a multiple scattering environment. In section II, we have presented numerical simulations and retrieved the Green's function (in phase and amplitude) between passive sensors. As a simple model, we chose to describe the coda as a superposition of waves arriving at random time. This offers the opportunity to develop, in sections III and IV, a simple estimation for the mean and the variance of the correlations, from which a SNR is derived. This SNR quantifies the convergence of the correlations toward the Green's function. The SNR was found to increase like $\sqrt{\frac{N\tau_D}{\tau_c}}$, where $N$ is the amount of sources, $\tau_{D}$ is the effective duration of the coda (the Thouless time), and $\tau_c$ represents the duration of the source. A central point of our paper is that multiple scattering (large $\tau_{\sigma}$ and $\tau_D$) provides a better SNR in the correlations. This results in a fascinating observation:  instead of blurring the images as is the case in conventional techniques, disorder here improves the quality of passive images. On the contrary, absorption was found in section V to reduce the SNR and, as always, is a limiting factor. Our theoretical model for the SNR was confronted to finite difference numerical simulations. This model was found to be valid when scattering is limited ($k\ell^{\star}>18$). Nevertheless, when scattering is increased, our naive description of coda waves was found to be slightly inappropriate: we point out short- and long-range correlations of diffuse wave paths as a probable candidate to explain the discrepancy between our theoretical model and numerical simulations.

\section*{ACKNOWLEDGEMENTS}
We are grateful to J. De~Rosny, L. Gizon, R. Maynard, L. Stehly, P. Gou\'edard and C. Hadziioannou for fruitful discussions and comments. The numerical code was provided by M. Tanter. This work was partially funded by an ANR grant "Chaire d'excellence-2005".

\appendix

\section{Variance of the correlations for one source}
To evaluate the level of fluctuations in the correlations, we evaluate the variance of the correlations for one source, averaged over the source position. Let's start with a simple calculation, assuming that $s(t)$ is gaussian and using the moment theorem:

\begin{widetext}
\begin{align}
&E\left\{\overline{C_{AB}(\tau)}^2\right\}=\iint E\left\{s_A(\theta_1)s_B(\theta_1+\tau)s_A(\theta_2)s_B(\theta_2+\tau)\right\}d\theta_1 d\theta_2\\
&= \iint E\left\{s_A(\theta_1)s_B(\theta_1+\tau)\right\}d\theta_1E\left\{ s_A(\theta_2)s_B(\theta_2+\tau)\right\}  d\theta_2\\
&+ \iint E\left\{s_A(\theta_1)s_A(\theta_2)\right\}E\left\{ s_B(\theta_1+\tau)s_B(\theta_2+\tau)\right\}  d\theta_2d\theta_1\\
&+ \iint E\left\{s_A(\theta_1)s_B(\theta_2+\tau)\right\}E\left\{ s_A(\theta_2)s_B(\theta_1+\tau)\right\}  d\theta_2d\theta_1\\
\end{align}

The first term is the intensity of the correlations. Once eliminated, only the variance remains:

\begin{align}
var \left\{\overline{C_{AB}(\tau)}\right\}=&\iint \sigma(\theta_1) \sigma(\theta_2)\sigma(\theta_1+\tau)\sigma(\theta_2+\tau) \left[\rho \otimes G_{AA}\right](\theta_1-\theta_2) \left[\rho \otimes G_{BB}\right](\theta_1-\theta_2)d\theta_2d\theta_1\\
+\iint& \sigma(\theta_1) \sigma(\theta_1+\tau)\sigma(\theta_2)\sigma(\theta_2+\tau) 
\left[\rho \otimes G_{AB}\right](\theta_1-\theta_2+\tau)\left[\rho \otimes G_{AB}\right](\theta_2-\theta_1+\tau)d\theta_2d\theta_1.
\end{align}

We replace $q$ by $\theta_2-\theta_1$:
\begin{align}
var \left\{\overline{C_{AB}(\tau)}\right\}=&\iint \sigma(\theta_1) \sigma(\theta_1+q)\sigma(\theta_1+\tau)\sigma(\theta_1+\tau+q)
 \left[\rho \otimes G_{AA}\right](q)\left[\rho \otimes G_{BB}\right](q)     dqd\theta_1 \\
+\iint& \sigma(\theta_1) \sigma(\theta_1+\tau+q)\sigma(\theta_1+q)\sigma(\theta_1+\tau) 
\left[\rho \otimes G_{AB}\right](\tau+q)\left[\rho \otimes G_{AB}\right](\tau-q)dqd\theta_1.
\end{align}
\end{widetext}

This central formula is similar to previous results by Derode et al in the case of TR \cite{derode1999}, and by Sabra et al. in the case of stationary noise correlations\cite{sabra2005a}. Nevertheless, we can go beyond their work after a series of additional reasonable assumptions.
1) In a dilute medium ($k\ell\gg 1$), $\left[\rho \otimes G_{AA}\right](q)$ and $\left[\rho \otimes G_{BB}\right](q)$ are constituted by the source autocorrelation (a peak around $q=0$)  followed by rapidly decaying contributions (reflections from surrounding heterogeneities). These contributions can be neglected: $G(t)\approx \delta(t)$, which implies $\left[\rho \otimes G_{AA}\right](q) \left[\rho \otimes G_{BB}\right](q)\approx \rho^2(q)$. 2) The variance $\sigma$ of the record evolves with a characteristic time of $\tau_{\sigma}$ much greater than the diffuse wave coherence time $\tau_c$ ($\tau_c\ll \tau_{\sigma}$), so that $\sigma(\theta+q)\approx \sigma(\theta)$ and $\sigma(\theta+q+\tau)\approx \sigma(\theta+\tau)$. 3) If $A-B\gg \lambda$, then $\left|G_{AB}\right|\ll\left|G_{AA}\right|$. These assumptions greatly simplify the above equation that now reads:

$$
var \left\{ \overline{C_{AB} (\tau)}\right\} \approx \int_0^T \sigma^2(\theta)\sigma^2(\theta+\tau) d\theta \int \rho^2(q)dq
$$

\section{Variance of the correlations for N sources}
We now take into account an averaging over $N$ sources, each source is labeled $i$ or $j$:
\begin{widetext}
\begin{align}
var\left\{\overline{C_{AB}(\tau)}\right\}=&\frac{1}{N^2}\sum_{ij} \iint E\left\{s_A^i(\theta_1)s_A^j(\theta_2)\right\}E\left\{ s_B^i(\theta_1+\tau)s_B^j(\theta_2+\tau)\right\}  d\theta_2d\theta_1\\
+&\frac{1}{N^2} \sum_{ij} \iint E\left\{s_A^i(\theta_1)s_B^j(\theta_2+\tau)\right\}E\left\{ s_A^j(\theta_2)s_B^i(\theta_1+\tau)\right\}  d\theta_2d\theta_1.
\end{align}
\end{widetext}

These summations can be split into two contributions $i=j$ and $i\neq j$. The first one yields $\frac{1}{N}\int_0^T \sigma^2(\theta)\sigma^2(\theta+\tau) d\theta \int \rho^2(q)dq$, and directly derives from to the above case N=1. The term $i\neq j$ 
exhibits cross-correlations $E\left\{s_A^i s_A^j\right\}$ and $E\left\{s_A^i s_B^j\right\}$, which are neglected as long as sources are dilute enough. This term $i\neq j$ contains short- and long-range correlations. The  short-range correlation in diffuse media takes the usual form of  $sinc(kd_{ij})e^{-d_{ij}/\ell^{\star}}$, with $d_{ij}$ the distance between source $i$ and $j$. This correlation ranges over one wavelength. Its role is developed in the next section of the appendix and can not account for the whole discrepancy between observed and theoretical SNR. When diffusion increases, their also exists some long-range contributions. These latter phenomena will be subject to further investigations.

\section{Spatial correlation of sources}\label{Neff}

We evaluate here the number of uncorrelated sources when $N$ sources are chosen at random but have are spatially correlated over $\lambda/2$, which is typically the case inside multiple scattering media (short-range correlation)~\cite{pnini1989}. We assume that the choice of $N$ locations is random over $M$ sites. The size of the site is deduced from the coherence length of a diffuse field, which result in a coherence area of a source of the form $\pi\lambda^2/4$. In the case of a $50 \lambda \times 50 \lambda$ large grid, we get $M\approx3183$ uncorrelated sites. The amount of uncorrelated sources $N_{eff}$ is therefore lower than $N$, as two sources can be chosen at the same site. We start with $N_{eff}(1)=1$. Then, we iterate:
$$
N_{eff}(N+1)=N_{eff}(N)+\frac{M-N_{eff}(N)}{M}.
$$

For the 3183 independent sites, we get $N_{eff}=1377$ independent sources, which result in lowering the SNR of 23\%. Short-range correlations can therfore not solely explain the discrepancies between the theoretical SNR and the SNR form numerical simulations. 




\end{document}